\documentclass[aps,floatfix,twocolumn,superscriptaddress,showpacs,preprintnumbers,longbibliography,nofootinbib]{revtex4-1}

\usepackage[utf8]{inputenc} %
\usepackage{amssymb,amsfonts,amsmath}
\usepackage{float} 
\usepackage{amsthm}
\usepackage{enumitem}
\usepackage{bbold}

\usepackage{hyperref} 
\hypersetup{colorlinks=true, urlcolor=blue, citecolor=blue,linkcolor=blue}
\usepackage[all]{hypcap} 

\usepackage{graphicx}
\usepackage{dcolumn}
\usepackage{bm}
\usepackage{psfrag}
\usepackage{epsfig}
\usepackage{color}
\usepackage{bbm}
\usepackage[FIGTOPCAP,raggedright,nooneline]{subfigure}
\usepackage{verbatim}
\usepackage{upgreek} 
\usepackage{physics}
\usepackage{tikz} 

\DeclareMathOperator*{\Motimes}{\text{\raisebox{0.25ex}{\scalebox{0.8}{$\bigotimes$}}}}

\newcommand{\ignore}[1]{}

\newcommand{\s}[2]{\sigma^{#1}_{#2}} 
 
\newcommand{\om}{\omega} 
\newcommand{\Om}{\Omega} 
 
\newcommand{\ad}{a^\dagger}
\newcommand{\bd}{b^\dagger} 
\newcommand{\dw}{\downarrow}
\newcommand{\up}{\uparrow}

\begin{document}

	\title{Reading a qubit quantum state with a quantum meter: time unfolding of quantum Darwinism and quantum information flux}
	
	\author{Salvatore Lorenzo}
	\affiliation{Dipartimento di Fisica e Chimica, Universit\`a  degli Studi di Palermo, via Archirafi 36, I-90123 Palermo, Italy}
	\author{Mauro Paternostro}
	\affiliation{Centre for Theoretical Atomic, Molecular, and Optical Physics, School of Mathematics and Physics, Queen's University, Belfast BT7 1NN, United Kingdom}
	\author{G. Massimo Palma}
	\affiliation{NEST, Istituto Nanoscienze-CNR and Dipartimento di Fisica e Chimica, Universit$\grave{a}$  degli Studi di Palermo, via Archirafi 36, I-90123 Palermo, Italy}
	
	\date{\today}
	
	\begin{abstract}
		Quantum non Markovianity and quantum Darwinism are two phenomena linked by a common theme: the flux of quantum information between a quantum system and the quantum environment it interacts with. In this work, making use of a quantum collision model, a formalism initiated by Sudarshan and his school, we will analyse the efficiency with which the information about a single qubit gained by a quantum harmonic oscillator, acting as a meter, is transferred to a bosonic environment. We will show how, in some regimes, such quantum information flux is inefficient, leading to the simultaneous emergence of non Markovian and non darwinistic behaviours
	\end{abstract}
	
	\maketitle
	
	\onecolumngrid

\section{Introduction}
George Sudarshan contributed in a significant way in forging the language and the mathematical tools which have now become  standard  in the description of open system dynamics, the introduction of the concept of dynamical map and a rigorous formulation of the Markovian master equation being just two of such contributions \cite{Sudarshan_61,Sudarshan_63}.  In recent years collision models have helped in clarifying the deep link between the two. Such repeated interaction models of open dynamics, to the best of our knowledge first introduced by Jayaseetha Rau \cite{Rau_63}, Sudarshan's PhD student, have proved to be powerful tools to derive mathematically correct forms of master equation beyond the Markovian regime. 
In this manuscript we will use a collision model to analyse some aspects of the flux of quantum information which takes places in the process of decoherence. 

Out of the many forms of irreversible dynamics, decoherence is probably the most relevant from the the viewpoint of fundamental physics. It can be described as the process by which the ability of a quantum system to show quantum interference is degraded due to its coupling with the environment. A paradigmatic example is the case of the dephasing of quantum spins in magnetic resonance, where following the BBP theory of relaxation \cite{BPP}, the environment is modelled as a classical random field describing the average fluctuating effect of the nearby atoms acting on the spin, inducing a time - dependent energy level splitting. 

It is however not yet completely clear which aspects of decoherence require a fully quantum description of the environment, and in particular which features are inextricably linked to the entanglement between the system and the environment degrees of freedom. 

Decoherence  transforms coherent quantum superposition into classical mixed states and 
plays a fundamental role in our understanding of the emergence of classicality in a quantum system as it
is now largely viewed as a process in which the environment measures the state of the quantum system it is coupled with
~\cite{zurek_environment-induced_1982,zurek_decoherence_2003,brandao_generic_2015,unden_revealing_2018,zwolak_complementarity_2013} .
The ensuing quantum information flow from the system to the environment is in turn the main focus of two timely research areas in the field of open system dynamics, namely non Markovianity~\cite{rivas_quantum_2014,breuer_colloquium:_2016,de_vega_dynamics_2017,apollaro_ meter-keeping_2011,breuer_foundations_2012,laine_measure_2010,breuer_measure_2009,rivas_entanglement_2010,lorenzo_geometrical_2013} and quantum Darwinism~\cite{zurek_quantum_2009,blume-kohout_simple_2005}.

Quantum Darwinism strives to explain the emergence of  classical reality out of quantum superpositions by looking at how the information the environment acquires about the system is accessible to separate observers~\cite{ollivier_objective_2004,blume-kohout_quantum_2006,lampo_objectivity_2017}. In this new paradigm the environment is assumed to consist of several independent fragments, whose degrees of freedom are not traced out but rather carry information about the system~\cite{zwolak_quantum_2009,zwolak_redundant_2010,zwolak_amplification_2014,zwolak_amplification_2016,zwolak_redundancy_2017,horodecki_quantum_2015,korbicz_objectivity_2014,riedel_rise_2012}. To be more specific, let us assume the system - environment interaction has a privileged pointer basis $|\pi_k\rangle$~\cite{zurek_pointer_1981,brunner_coupling-induced_2008} and that, as a consequence of such interaction, a coherent superposition of pointer states
$|\Psi\rangle_S = \sum_{k=1}^n \psi_k |\pi_k\rangle_S$ evolves into a joint system-environment state with a branching structure  
\begin{equation}
|\Psi_{SE}\rangle = \sum_{k=1}^n \psi_k |\pi_k\rangle_S\bigotimes^M_{j=1}|\eta_k\rangle_j.
\label{branch}
\end{equation}
In Eq. (\ref{branch})  the information about the system state $|\pi_k\rangle_S$ is imprinted into multiple copies of environmental states $|\eta_k\rangle$. The above state clearly shows a multipartite GHZ - like entangled structure. If separate observers can access each a separate environmental fragment, i.e. each can measure a separate set of environmental degrees of freedom, and if the amount of information each fragment contains about the system is the same, we are in the presence of redundant information about the system and separate observer all agree on the system state. In other words, as we will elaborate later, the signature of quantum Darwinism is the presence of a redundancy plateau in the mutual information between the system and environmental fragment as a function of the fragment size.

The issue of how quantum information flows from the system to the environment is a key element in our understanding of
Quantum non-Markovianity.
The advent of quantum information theory has provided in recent years the mathematical tools to understand and characterise non-Markovian dynamics in the quantum dynamics scenario~\cite{de_vega_dynamics_2017,apollaro_ meter-keeping_2011}. In particular, a number of theoretical measures of the degree of quantum non-Markovianity of an open dynamics have been put forward~\cite{breuer_foundations_2012,laine_measure_2010,breuer_measure_2009,rivas_entanglement_2010,lorenzo_geometrical_2013}.
All such measures capture the idea that non-Markovianity is linked to a back-flow of quantum information from the environment to the system. As far as decoherence is concerned all such measure lead to the same criterion to discriminate between Markovian and non Markovian quantum dynamics (this is not in general te case for other instances of irreversible quantum dynamics). 

Several papers have recently analysed the link between quantum information backflow and quantum Darwinism, which, as said  before, is strongly linked with the process of quantum decoherence. 
In particular the detrimental role of non Markovianity on the emergence of Darwinism has been shown in~\cite{galve_non-Markovianity_2016,giorgi_quantum_2015,pleasance_application_2017}. Similar conclusions have been reached in~\cite{milazzo}, where the assumption of  interacting environmental fragments has been shown to lead to the existence of quantum information trapping, with an ensuing  sudden transition between Markovian/Darwinistic to non Markovian/non Darwinistic behaviour. 
The time unfolding of quantum Darwinism has been studied in \cite{lorenzo_2019}, where its engineering via Zeno/anti-Zeno effect has been proposed. 
Again a random collision model \cite{perez_2019} has shown that, in the absence of system-environment entanglement one can observe non Markovianity but not Darwinism. 

In this manuscript we will analyse  the situation in which the environment does not acquire information via a direct interaction with the system , but rather it reads it via a meter coupled with the system. Mediated system - environment couplings are known to lead to non trivial effects, ranging from non Markovian dynamics~\cite{ciccarello_NM,ciccarello_NM2} to inefficient information erasure~\cite{lorenzo_2015}.
In the following we will assume that, by entangling itself with the system pointer states of the system, the meter acquires information on the system state. Such information is then transferred to the environmental degrees of freedom, i.e. the fragments external observer can access, learning indirectly about the system state.
The manuscript is structured as follows: in the following section we will introduce the system and we will analyse its degree of non Markovian dynamics.
We will then analyse the degree of Darwinism of the process. We will end with our conclusions

\section{A collision model analysis of Non-Markovian indirect decoherence}

The model consists of a two-level system $\mathcal{S}$, with gap energy $\varepsilon$, whose ground state $\ket{\dw}$ and excited state $\ket{\up}$ are coupled dispersively with a harmonic oscillator $\mathcal{M}$, of frequency $\omega_{\mathcal{M}}$, acting as a meter,
in turn coupled to a quantum environment consisting of an infinite number of oscillators $\mathcal{R}$. We will assume the following system + meter + reservoir hamiltonian:
\begin{eqnarray}
H{=}\varepsilon\s{z}{\mathcal{S}}+\om_\mathcal{M} \ad a+\int \!\! d\om\; \om\, \bd(\om)b(\om)+
\Om \s{z}{\mathcal{S}}(\ad+a)+\int\!\! d\om \;\kappa(\om)\left(\ad b(\om)+a \bd(\om)\right)\nonumber
\label{H}
\end{eqnarray}
where $\sigma^{z}{=}\ket{\up}_\mathcal{S}\!\bra{\up}{-}\ket{\dw}_\mathcal{S}\!\bra{\dw}$ is the usual pseudo-spin operator while $a$ and $b(\om)$ are the annihilation operators of $\mathcal{M}$ and $\mathcal{R}$ respectively which satisfy the standard bosonic commutation relations $[a,a^\dagger]=1$ and $[b(\om),b(\om')^\dagger]{=}\delta(\om-\om')$.

It is rather straightforward to get a physical intuition of the dynamics generated by (\ref{H}): the qubit - meter interaction induces mutual entanglement between the two, with the meter ${\mathcal M}$ reading the state of the qubit. Such quantum information is then transferred to the environmental degrees of freedom. An effective picture consists of a qubit coupled to a cavity mode, with the latter coupled to travelling field modes. The output field modes thus contain information on the qubit which can be accessed by external independent observers. 
The above picture clearly calls for an input-output approach, which has been shown recently to be equivalent to a collision model
\cite{ciccarello_QMQM_2017}, a convenient picture as we are interested in the time unfolding of the quantum information flow. Furthermore such models have the clear advantage to allow for an unambiguous identification of the accessible environmental fragments. 

To map the above dynamics into a collision model let us first move to the interaction picture with respect to the free Hamiltonian 
\begin{eqnarray}
H_{I}(t)=\Om \s{z}{S}(\ad e^{i\om_M t}+a e^{-i\om_M t})+
\sqrt{\frac{\gamma}{2\pi}}\int d\om \left(\ad b(\om)e^{-i \Delta t}+a \bd(\om)e^{i\Delta t}\right)
\label{Hi}
\end{eqnarray}
where we assumed a constant coupling between the  meter and its reservoir $\kappa(\om){=}\sqrt{\gamma/2\pi}$ and $\Delta{=}\om{-}\om_M$.
Defining the quantum white noise operators \cite{ciccarello_QMQM_2017} $b(t)$ as
\begin{eqnarray}
b(t)=\frac{1}{\sqrt{2 \pi}}\int\!\! d\om\, b(\om)e^{-i\Delta t}
\label{input}\end{eqnarray}
we rewrite (\ref{Hi}) as
\begin{eqnarray}
H_{I}(t)=\Om \s{z}{S}(\ad e^{i\om_M t}{+}a e^{-i\om_M t})+\sqrt{\gamma}\left(\ad b(t){+}a \bd(t)\right)\nonumber
\label{H2}\end{eqnarray}
The time evolution operator generated by $H_I(t)$ can be formally written as
\begin{equation}
U(t)=\prod_{k=1}^{n}U_k=\prod_{k=1}^n \mathcal{T}e^{-i \int_{(k{-}1)\tau}^{k\tau} H_I(t')dt'}
\end{equation}
where we have divided time in steps of duration $\tau$.
In the spirit of collision models, following \cite{ciccarello_QMQM_2017}, we now  assume $H_I(t){=}H_k$ constant in each $k$-th interval, and equal to its time average value, i.e.
\begin{equation}
H_k= \Om \s{z}{S}(\ad f^*_k+a f_k)+\sqrt{\gamma /\tau}\left(\ad b_k+a \bd_k\right)
\end{equation}
where 
\begin{eqnarray}
f_k{=}\frac{1}{\tau}\int_{(k{-}1)\tau}^{k\tau}\!\!\!e^{-i\om_M t'}dt'
\;\;\text{and}\;\;
b_k{=}\frac{1}{\sqrt{\tau}}\int_{(k{-}1)\tau}^{k\tau}\!\!\!b(t')dt'.
\label{discrete}
\end{eqnarray}
Note that the $\sqrt{\tau}$ factor in \eqref{discrete} guarantees correct bosonic commutation relations for the $b_k$ operators \cite{ciccarello_QMQM_2017}.

By making use of a Suzuki-Trotter approach we expand each $U_k$, neglecting terms $O(\tau^2)$. The dynamics then reduces to a sequence of pairwise system - meter and meter - environment collisions
\begin{equation}
U_k=U^\mathcal{SM}_k U^\mathcal{MR}_k=e^{\s{z}{S}(\ad\alpha_k{-}a \alpha_k^*)}e^{i\theta/2(\ad b_k{-}a \bd_k)}
\end{equation}
where 
\begin{eqnarray}
\alpha_k=-i\Om \tau f_k \quad\text{and}\quad
\theta=-2\sqrt{\gamma \tau}
\end{eqnarray}
Note that $U^\mathcal{SM}$ takes the form of a conditional displacement operator dependent on the state of the two-level system,
while  $U^\mathcal{MR}$ acts as a beam-splitter. We stress again the role of the time - dependent reservoir modes $k$ in the above picture: they describe the state of the environment which interacts with the meter at time $k\tau$. In other words, they play the role of the ancillae in standard collision models. Note that, after the collision the $k$ modes carry information on the system, which is immediate to see by observing that the initial state 
\begin{eqnarray}
\ket{\psi^0}=(a\ket{\dw}_{\!\mathcal{S}}+\beta\ket{\up}_{\!\mathcal{S}})\Motimes\ket{0}_{\!\mathcal{M}}\Motimes_{j=1}^{n}\ket{0}_{\!\mathcal{R}_j}
\end{eqnarray}
will evolve,  after $\ell$ collision, into
\begin{eqnarray}
\ket{\psi^\ell}
{=}\left[\alpha\ket{\dw}_{\!\mathcal{S}}\Motimes\ket{\Om_-^\ell}_{\!\mathcal{M}}\Motimes_{j=1}^{\ell}\ket{\gamma_+^j}_{\!\mathcal{R}_j}+\beta\ket{\up}_{\!\mathcal{S}}\otimes\ket{\Om_+^\ell}_{\!\mathcal{M}}\Motimes_{j=1}^{\ell}\ket{\gamma_-^j}_{\!\mathcal{R}_j}\right]
\Motimes_{j=\ell+1}^{n}\ket{0}_{\!\mathcal{R}_j}\nonumber
\end{eqnarray}
where the  meter and the environmental ancillae are in coherent states of amplitude respectively
\begin{eqnarray*}
	\Om^\ell_\pm=(\Om^{\ell{-}1}_\pm{\pm}\alpha_{\ell})\cos(\theta/2),\\
	\gamma^\ell_\pm=i(\Om^{\ell{-}1}_\pm{\pm}\alpha_{\ell})\sin(\theta/2).
\end{eqnarray*}
The usual first effect of such system environment entanglement is decoherence. Indeed the reduced density matrix of the system after $\ell$ collisions reads  
\begin{eqnarray}
\rho_{\mathcal{S}}^\ell=\tr_\mathcal{R}\{\ket{\Psi_\ell}\bra{\Psi_\ell}\}=\left(\begin{array}{cc}   |\alpha|^2 & \alpha\beta^*\kappa^\ell\\ \alpha^*\beta\kappa^{*\ell} & |\beta|^2\end{array}\right),
\label{rhoS}
\end{eqnarray}
which shows clearly how the off diagonal matrix elements shrink as $\kappa^{(\ell )}$, where the step dependent factor $\kappa^{(\ell )}$ is the product of two terms, the first due to the meter and the second to the ancillae
\begin{eqnarray}
\kappa^{(\ell )}=\braket{\Om_+^\ell}{\Om_-^\ell}_{\mathcal{M}}\prod_{j=1}^{\ell}\braket{\gamma_+^j}{\gamma_-^j}_{\mathcal{R}_j}.
\label{ks}\end{eqnarray}
In terms of the internal product between two coherent states, noting that $\Om_-^\ell=-\Om_+^\ell$, we have 
\begin{eqnarray}
\kappa^{(\ell )}_\mathcal{M} &=&\braket{\Om_+^\ell}{\Om_-^\ell}_{\mathcal{M}}=e^{-2|\Om^\ell_+|^2},\\
\kappa^{(\ell )}_\mathcal{R}&=&\prod_{j=1}^{\ell}\kappa^{(j)}_\mathcal{R}=\prod_{j=1}^{\ell}\braket{\gamma_+^j}{\gamma_-^j}_{\mathcal{R}_j}=e^{\sum_{j=1}^\ell 2|\gamma_+^j|^2}.
\end{eqnarray}
The decoherence process is better analysed in terms of rates. To this end let us define $\Gamma (\ell ){=} \Gamma_{\mathcal{R}}(\ell ){ + }\Gamma_{\mathcal{M}}(\ell )$
with
$\Gamma_{\mathcal{M}} (\ell ){ = -}\ln \kappa_{\mathcal{M}}^{(\ell )}$ and analogously $\Gamma_{\mathcal{R}} (\ell ){ = -}\ln \kappa_{\mathcal{R}}^{(\ell )}$. 
\begin{figure}
	\centering
	\includegraphics[width=0.85\linewidth]{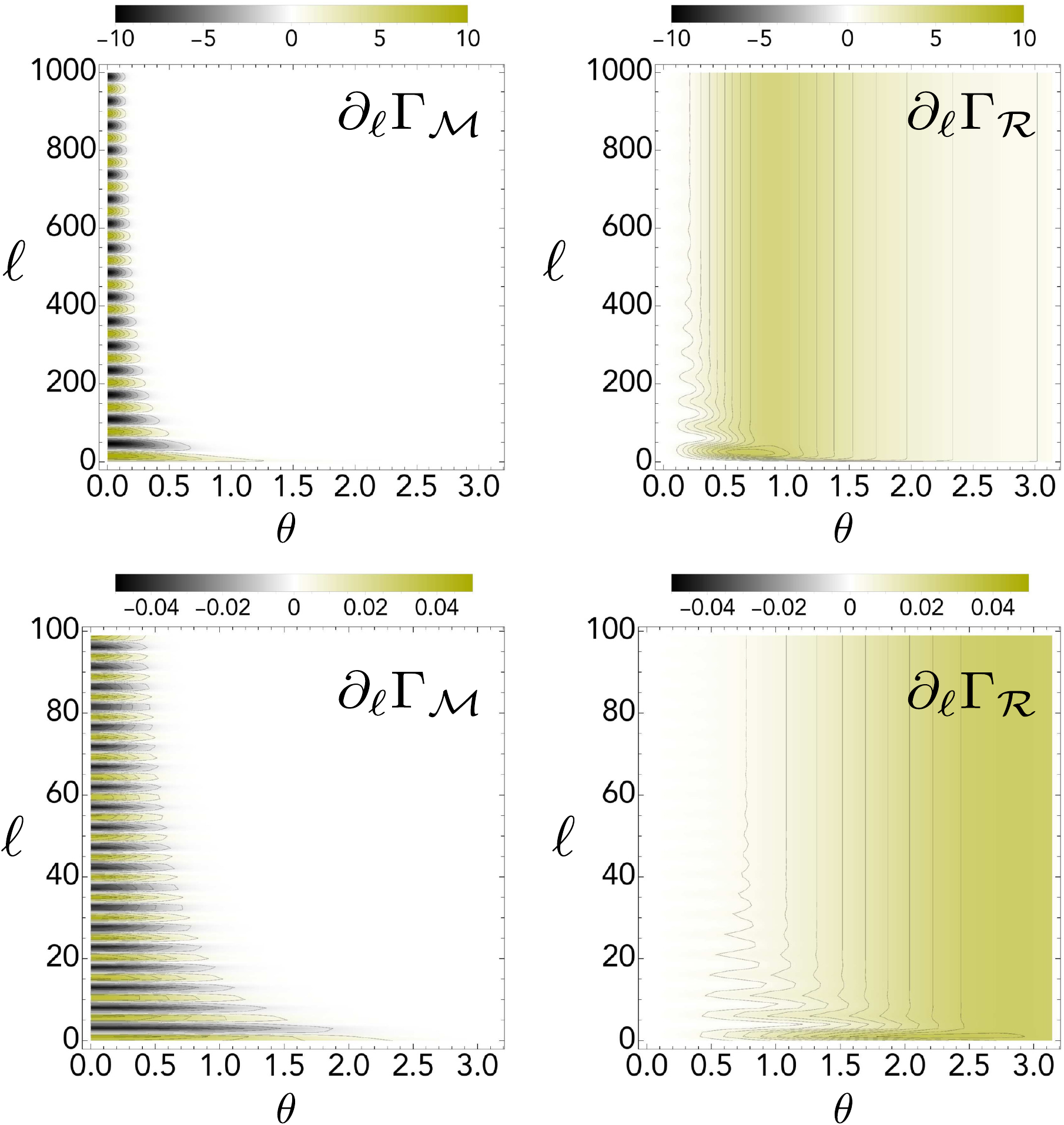}
	\caption{Decay rates as function of time (the number of collisions $\ell$), for $\Omega{=}0.5$ and $\omega_M{=}0.1$ (top), and $\omega_M{=}5$ (bottom). The contribution of the  meter to the decay rate of the system are shown on the left while on the right the contribution of  ancillae are shown. Note that only the  meter, with its oscillations for small values of $\theta$ is responsible of possible non-Markovian behaviour of the system dynamics}
	\label{fig:rates}
\end{figure}
As mentioned in the introduction, even if the meaning of non-Markovianity can be explained through very different interpretations\cite{breuer_measure_2009,rivas_entanglement_2010,lorenzo_geometrical_2013} for the reduced dynamics under study in this work, all these different interpretations converge to the same result. Indeed the characterisation of non Markovianity of the system dynamics for pure decoherence is straightforward as all the main  non Markovianity measures agree on the fact that decoherence is non Markovian whenever  $ \partial_{\ell}\Gamma (\ell ) $ is negative. In Fig(\ref{fig:rates}) one can immediately see that, while $\partial_{\ell}\Gamma_{\mathcal{R}} (\ell )$ is always positive, $ \partial_{\ell}\Gamma_{\mathcal{M}}(\ell ) $ oscillates between positive and negative values for small values of $\theta$, signature of the presence of a non Markovian dynamics. Such inefficient quantum information transfer between system and environment mediated by the quantum meter is more pronounced for stiffer meters as 
$ \partial_{\ell}\Gamma_{\mathcal{M}}(\ell ) $ shows stronger and more rapid oscillations
for larger values of $\omega_{\mathcal{M}}$. One can expect that in this regime part of the quantum information about the system remains trapped in the meter. We will show that this is indeed the case.


\section{Quantum Darwinism vs quantum non Markovianity}\label{Darw}
The degree of Darwinism of the reduced system dynamics is quantified in terms of the mutual information shared by an environmental fragment consisting of a given number collided ancillae and the system. The number of times that this information can be independently extracted is taken as measure of objectivity. 
The mutual information between the system and a subset $m$ of  ancillae after $\ell$ collisions
\begin{eqnarray}
\mathcal{I}(\mathcal{S},\mathcal{F}^\ell_m)=
S(\rho_{\mathcal{S}}^{\ell})+S(\rho_{\mathcal{F}_m}^\ell)-S(\rho_{\mathcal{SF}_m}^\ell)
\end{eqnarray}
is the information about $\mathcal{S}$ available from $\mathcal{F}_m$. We call ${\cal F}^\ell_{\delta}$ the smallest fraction of the environment such that, for any arbitrarily picked information gap $\delta$, we achieve 
\begin{eqnarray}
\mathcal{I}(\mathcal{S},\mathcal{F}^\ell_{\delta})\geq(1-\delta)S(\rho_{\mathcal{S}}^{\ell}).
\label{deficit}
\end{eqnarray}
We can thus define the redundancy as 
\begin{eqnarray}
R{=}\ell/\mathcal{F}^\ell_\delta,
\end{eqnarray}
which provides a quantitative estimate of the number of distinct subenvironments that supply classical information about ${\cal S}$, up to the deficit $\delta$.

To evaluate the mutual information between the system and growing environmental fragments, we make extensive use of the fact that the overall system-meter-environment state remains pure after an arbitrary number of collisions.
The von Neumann entropy of $\mathcal{S}$ is easily calculated at each step as
\begin{eqnarray}
S(\rho_{\mathcal{S}}^\ell)=-\lambda_\mathcal{S}^+\log_2\lambda_\mathcal{S}^+-\lambda_\mathcal{S}^-\log_2\lambda_\mathcal{S}^-
\label{VNE}\end{eqnarray}
with $\lambda^\pm_\mathcal{S}=\frac{1}{2}\left(1\pm\sqrt{1-4(1-|k^{\ell}|^2)|\alpha|^2|\beta|^2}\right)$.
From now we omit the Hilbert space pedices.
Let us consider a sequence of $m$ ancillae. We observe that, after $\ell>m$ collisions, the reduced density matrix of the fragment $\mathcal{F}_m$ is given by
\begin{eqnarray}
\rho_{\mathcal{F}_m}^\ell=
|\alpha|^2\Motimes_{j=1}^{m}\;\ketbra{\gamma_+^j}{\gamma_+^j}{+}
|\beta|^2\Motimes_{j=1}^{m}\;\ketbra{\gamma_-^j}{\gamma_-^j},
\label{rhoF}
\end{eqnarray}
which implies that after the first $m$ collisions the reduced density matrix of $\mathcal{F}_m$ doesn't change anymore.
The same is not true for reduced state of the joint system $\mathcal{S}{-}\mathcal{F}_m$ due to the presence of the  meter $\mathcal{M}$, indeed we have
\begin{eqnarray}
\rho^\ell_{\mathcal{S}\mathcal{F}_m}{=}
|\alpha|^2 \Motimes_{j=1}^{m}\ketbra{\gamma_+^j}{\gamma_+^j}+ \alpha\beta^*\kappa_\mathcal{M}^\ell\Motimes_{j=1}^{m}\ketbra{\gamma_+^j}{\gamma_-^j}
{+}\alpha^*\beta\kappa_\mathcal{M}^{\ell*}\Motimes_{j=1}^{m}\ketbra{\gamma_-^j}{\gamma_+^j}+
|\beta|^2\Motimes_{j=1}^{m}\ketbra{\gamma_-^j}{\gamma_-^j}\nonumber
\label{rhoSF}
\end{eqnarray}
Owing to the fact that at $\ell$-th step the state of $\mathcal{SMR}_{\{1\dots \ell\}}$ remains a pure state, we can evaluate the Von Neumann entropy of joint system $\mathcal{S{-}F}_m$ looking at the the reduced density matrix of $\mathcal{M}$ and all the  ancillae already involved in a collision but out of $\mathcal{F}_m$, i.e. $\mathcal{F}_m^\perp\equiv\mathcal{R}_{\{m{+}1\dots \ell\}}$
\begin{eqnarray}
\rho^\ell_{\mathcal{MF}_m^\perp}=
|\alpha|^2\ketbra{\Om_+^\ell}{\Om_+^\ell}\Motimes_{j=m+1}^{\ell}\ketbra{\gamma_+^j}{\gamma_+^j}
+ |\beta|^2\ketbra{\Om_-^\ell}{\Om_-^\ell}\Motimes_{j=m+1}^{\ell}\ketbra{\gamma_-^j}{\gamma_-^j} \nonumber  
\end{eqnarray}
Defining $\ket{\Om_\perp^\ell}$ and $\ket{\gamma_\perp^j}$ as the orthogonal states to $\ket{\Om_+^\ell}$ and $\ket{\gamma_+^j}$ respectively
\begin{eqnarray}
\ket{\Om_\perp^\ell}=\frac{\ket{\Om_-^\ell}-\kappa^\ell_\mathcal{M}\ket{\Om_+^\ell}}{\sqrt{1-(\kappa^\ell_\mathcal{M})^2}}
\quad \text{and}\quad
\ket{\gamma_\perp^j}=\frac{\ket{\gamma_-^j}-\kappa^\ell_\mathcal{R}\ket{\gamma_+^j}}{\sqrt{1-(\kappa^j_\mathcal{R})^2}}\nonumber
\end{eqnarray}
we can write $\rho^\ell_{\mathcal{MF}_m^\perp}$ in the basis $\{\ket{\Om_p^\ell}\Motimes_{j}\ket{\gamma_q^j}\}$.
It is possible then possible to calculate the Von Neumann entropy of $\rho^\ell_{\mathcal{MF}_m^\perp}$ in terms of its eigenvalues 
\begin{equation}
\lambda_{\mathcal{MF}_m^\perp}^{\pm}{=}1/2(1{\pm}\sqrt{1-4|\alpha|^2|\beta|^2(1-|\kappa_\mathcal{M}^\ell|^2|\kappa_{\mathcal{F}_m^\perp}^\ell|^2)})
\end{equation}.
In a similar way it is possible to evaluate the vvonon Neumann entropy of $\rho_{\mathcal{F}_m}^\ell$ through its eigenvalues 
\begin{equation}
\lambda_{\mathcal{F}_m}^{\pm}{=}1/2(1{\pm}\sqrt{1-4|\alpha|^2|\beta|^2(1-|\kappa_{\mathcal{F}_m}^\ell|^2)})
\end{equation}
\begin{figure}
	\includegraphics[width=0.95\linewidth]{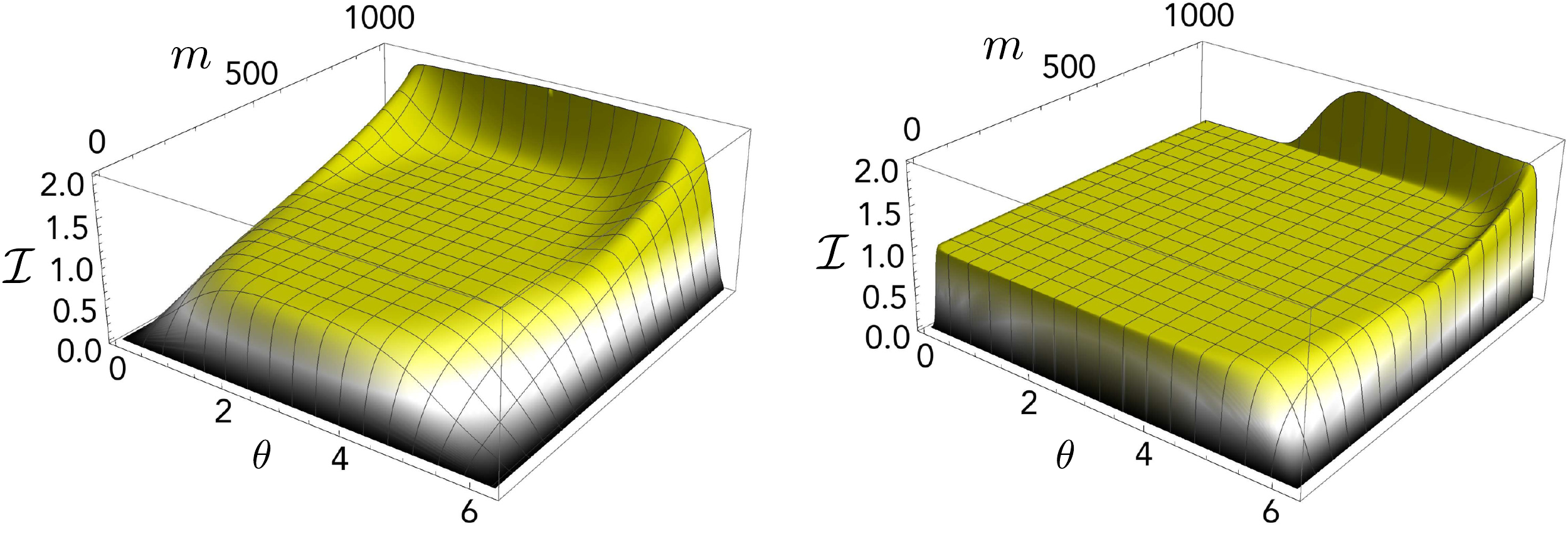}
\caption{Mutual information of $\rho^\ell_{\mathcal{SF}_m}$, for an initially balanced superposition of system poiner states,  as function of $m$ after $\ell=1000$ collisions for different values of $\omega_\mathcal{M}$. For $\omega_\mathcal{M}=0.1$ (left panel) the characteristic plateau typical of the emergence of quantum Darwinism is present for any values of $\theta$. Interestingly for small values of $\theta$ part of information about the system remain trapped in the meter as the mutual information doesn't rise to $2$. For $\omega_\mathcal{M}=5.0$ (right panel), a plateau emerges only for $\theta\sim 3$ and with a much smaller redundancy. }
	\label{fig:mi}
\end{figure}
The emergence of an objective reality
manifests itself with a plateau in the plot of $\mathcal{I}^\ell(\mathcal{S}:\mathcal{F}_m)$ as a function of the fragment size. In this case once the values of $\mathcal{I}^{\tilde\ell}(\mathcal{S}:\mathcal{F}_{\tilde m})$ for a fixed $\tilde\ell$
reaches the value of the system's entropy, a small portion $\tilde m$ of the
environment contains almost all information classically accessible on the system: enlarging the size of the fragment does not increase the amount of information accessible on the system and this information is redundantly stored in the environment. Defining an information deficit $\delta$, the redundancy $R^{\tilde \ell}_{\delta}=\tilde\ell/\tilde m$ is the number of independent fragments of the environment that supply almost all classical information on the system, i.e.
$(1-\delta) S(\rho_\mathcal{S}^{\tilde\ell})$. Large redundancy means that many observers can found independently the state of the system. 
Fig.(\ref{fig:mi}) shows clearly the dependence of the mutual information between environment and system on the stiffness of the meter.
For a soft meter ($\omega_\mathcal{M}=0.1$) the mutual information plateau characteristic of the emergence of quantum Darwinism is
present for all values of $\theta$. Note that for small values of $\theta$, when $\partial_{\ell}\Gamma_{\mathcal{M}} $ shows ascillations, some quantum information about the system remains trapped in the meter and the mutual information does not reach the value $\mathcal{I}^\ell(\mathcal{S}:\mathcal{F}_m)=2$ for $m\sim\ell$ typical of a system - environment pure state.
A different behaviour appears for a stiffer meter, as shown in (\ref{fig:mi}) for ($\omega_\mathcal{M}=5$). In this case the darwinistic plateau appears only for $\theta\sim 3$ and furthermore with a much smaller redundancy.

\begin{figure}
	\centering
	\includegraphics[width=0.95\linewidth]{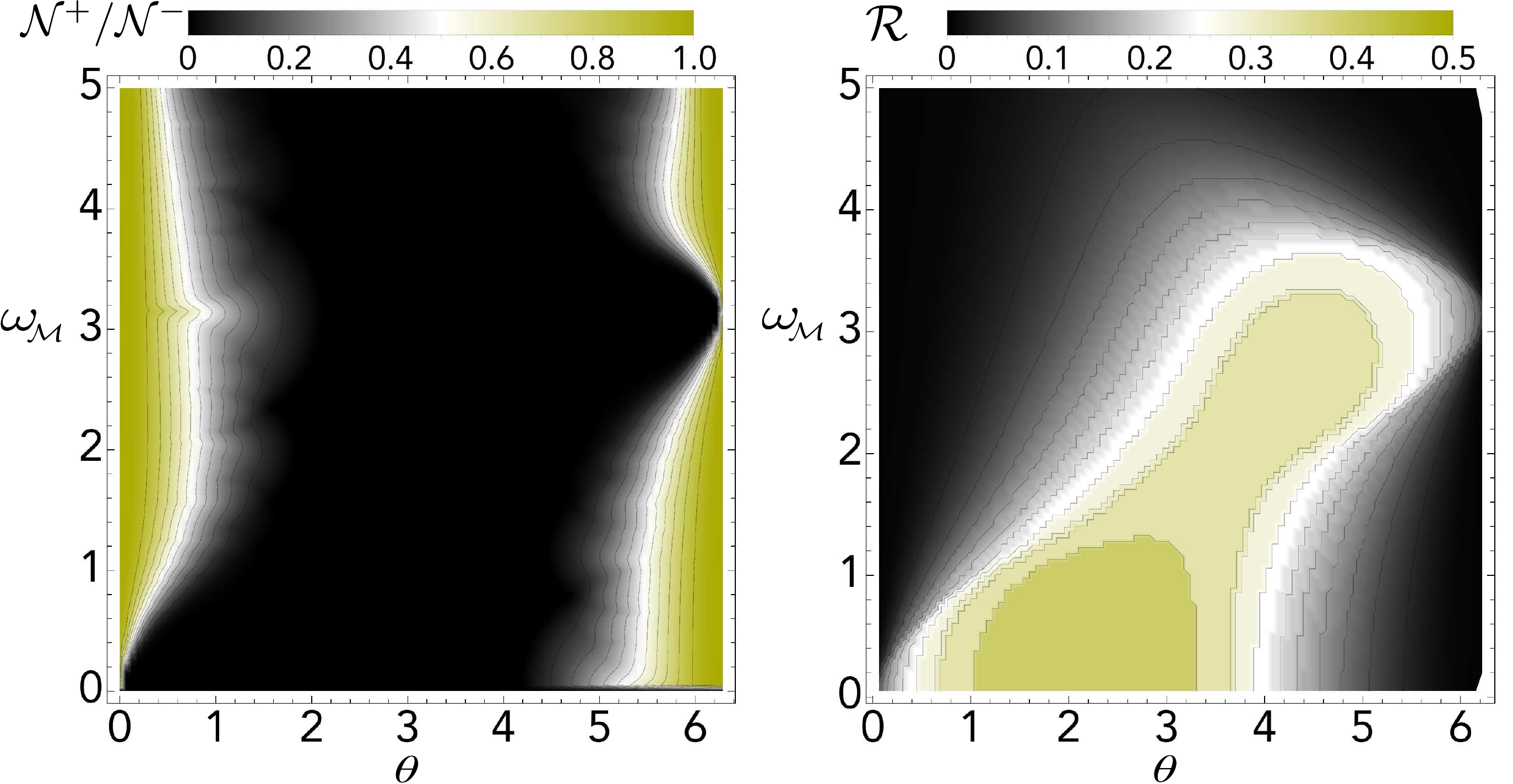}
	\caption{Normalized Non-Markovianity measure /left) as function of $\omega_\mathcal{M}$ and $\theta$ for $\Omega=0.1$. Redundancy $\mathcal{R}$ evaluated with an information gap, cfr \ref{deficit}, $\delta=0.1$ (right)}
	\label{fig:NM_R} 
\end{figure}

The two examples above seem to suggest that also for our model of quantum information flux mediated by a quantum meter the regime
in which the system deviates from Darwinism is also a regime characterised by quantum non Markovianity.
To analyse whether this is indeed the case we need to use a specyfic quantitative measure of non Markovianity. A convenient measure is the one introduced in \cite{lorenzo_geometrical_2013}. This is related to the changes in time of the volume of accessible states of the open system $V(t)$. Given that this volume can only decrease with time for a Markovian dynamics, the amount of quantum non-Markovianity is measured by
\begin{equation}
\mathcal{N}_V=\frac{1}{V(0)}\int_{\partial_t{V(t)}>0}\!{\rm d}t\,\,\partial_t{V(t)}\,,
\label{NM_nostra}
\end{equation}
where the integral is over the time domains in which $V(t)$ increases. 
Here we will adopt a more convenient normalization quantifing the degree of non Markovianity in terms of 
$ \frac{{\cal N}^+(\tau)}{{\cal N}^-(\tau)}$, where $ {\cal N}^+(\tau)$ (${\cal N}^-(\tau)$)  is defined  as plus (minus) the integral of $\partial_t V(t)$, over the time intervals in which it is positive (negative), but only up to fixed time $\tau$.
In Fig.(\ref{fig:NM_R}) the degree of normalized non Markovianity and the redundancy are plotted as a function of $\theta$ and $\omega_\mathcal{M}$. It emerges clearly that one reaches the highest redundancy when the dynamics is fully Markovian and that when the dynamics is non Markovian the redundancy is basically equal to zero. However at variance with \cite{milazzo} here we do not observe any sharp threshold between Markovianity and Darwinism to non Markovianity and non Darwinsism. For instance for a stiff meter one can still observe a Markovian dynamics and at the same time a very small redundancy. The reason for such behaviour is the indirect quantum information flux from the system to the environment through the meter.

\section{conclusions}
In our work we have again found evidence that whenever the system reduced dynamics is characterised by a non Markovian dynamics, Darwinism either does not emerge or it is charactyerized by a low degree of redundancy. The way in which this parallel features arise is dependent on the details of the model analysed. For instance both in \cite{milazzo} and in the present work we observe an inefficient quantum information flow from the system to the environment but while in \cite{milazzo} this is due to a spatial trapping of quantum information near the system leading to a sudden transition from a Markovian  and Darwinistic regime to to a non Markovian and non Darwinstic one, in the scenario analysed in the present paper, although part of the quantum information about the system remains trapped in time in the meter we observe a more nuanced transition from one regime to the other

\section*{acknowledgments}
 We  acknowledge support under PRIN project 2017SRNBRK QUSHIP funded by MIUR,  the EU Collaborative project TEQ (grant agreement 766900), the DfE-SFI Investigator Programme (grant 15/IA/2864), COST Action CA15220, the Royal Society Wolfson Research Fellowship ((RSWF\textbackslash R3\textbackslash183013), the Leverhulme Trust Research Project Grant (grant nr.~RGP-2018-266). G.M.P would like to thank the Kavli Institute for  Theoretical Physics, UC Santa Barbara, for its hospitality supported in part by the National Science Foundation under Grant No. NSF PHY-1748958.

\end{document}